# Antiferromagnetic skyrmion crystal in Janus monolayer CrSi$_2$N$_2$As$_2$

Kaiying Dou, Wenhui Du, Zhonglin He, Ying Dai*, Baibiao Huang, and Yandong Ma*

School of Physics, State Key Laboratory of Crystal Materials, Shandong University, Shandanan Street 27, Jinan 250100, China

*Corresponding author: daiy60@sina.com (Y.D.); yandong.ma@sdu.edu.cn (Y.M.)

**Abstract**

Antiferromagnetic skyrmion crystal (AF-SkX), a regular array of antiferromagnetic skyrmions, is a fundamental phenomenon in the field of condensed-matter physics. So far, only very few proposals have been made to realize the AF-SkX, and most based on three-dimensional (3D) materials. Herein, using first-principles calculations and Monte-Carlo simulations, we report the identification of AF-SkX in two-dimensional lattice of Janus monolayer CrSi$_2$N$_2$As$_2$. Arising from the broken inversion symmetry and strong spin-orbit coupling, large Dzyaloshinskii–Moriya interaction is obtained in Janus monolayer CrSi$_2$N$_2$As$_2$. This, combined with the geometric frustration of its triangular lattice, gives rise to the skyrmion physics and long-sought AF-SkX in the presence of external magnetic field. More intriguingly, this system presents two different antiferromagnetic skyrmion phases, and such phenomenon is distinct from those reported in 3D systems. Furthermore, by contacting with Sc$_2$CO$_2$, the creation and annihilation of AF-SkX in Janus monolayer CrSi$_2$N$_2$As$_2$ can be achieved through ferroelectricity. These findings greatly enrich the research on antiferromagnetic skyrmions.

**Keywords:** Antiferromagnetic skyrmion crystal, two-dimensional lattice, first-principles, Dzyalohinskii–Moriya interaction, Janus monolayer

**Introduction**

Magnetic skyrmions are localized vortex or antivortex like spin structures characterized by spins wrapping a unit sphere and protected by an integer topological charge Q [1-4]. Their topological properties lead to the exceptional inherent stability against transitions into trivial spin structures, making them technologically promising for data-storage applications [5-9]. On a fundamental level, magnetic skyrmions emphasize the role of topology in condensed matter physics [3,4]. In most systems observed, noncoplanarity in the directions of spins at different lattice sites correlates to ferromagnetic (FM) exchange interaction, namely FM skyrmions [10-19]. Nonetheless, FM skyrmions involve shortcomings in common with other spintronics, for example, the sensitivity to stray field [3,20,21]. By contrast, topological spin structures in antiferromagnetic (AFM) lattices, consisting of compensated FM sublattices, generate no stary field; and moreover, they are demonstrated to host ultrafast spin dynamics and relativistic motion of textures [22-28]. Therefore, generating AFM skyrmions is crucial for both fundamental research and applications [22, 29-35].

Despite the huge interest, AFM skyrmions have been experimentally observed only in several thin film systems, including $Fe_2O_3$-Pt [36], $MnSc_2S_4$ [37] and synthetic antiferromagnets [38-40]. In these few existing systems, the isolated AFM skyrmions are usually observed, while the AFM skyrmion crystal (AF-SkX), regular array of antiferromagnetic skyrmions, is rarely identified [37]. With the recent rise of long-range magnetic ordering in two-dimensional (2D) lattices, atomically thin AFM materials has gained a great deal of attention [41-44]. Physically, triggering by competing exchange interactions, 2D AFM systems can manifest topological spin textures as well, providing a new category of AFM skyrmion medium [30-35]. However, at present, few 2D AFM systems are demonstrated to be with skyrmion physics [45-47]. Actually, up to now, AF-SkX has not yet been realized in natural 2D materials [48].

In this work, we show that the AF-SkX can be realized in 2D lattice of Janus monolayer $CrSi_2N_2As_2$ based on first-principles calculations and Monte-Carlo (MC) simulations. Janus monolayer $CrSi_2N_2As_2$ is demonstrated to exhibit large Dzyaloshinskii–Moriya interaction (DMI) due to the joint effect of the broken inversion symmetry and strong spin-orbit coupling (SOC). This along with the geometric frustration of its triangular lattice stabilizes the skyrmion physics and AF-SkX under external magnetic field. With introducing skyrmion number Q and total chirality $\chi_L$ as measurements, the evaluation of AFM topological spin textures in Janus monolayer $CrSi_2N_2As_2$ with the magnetic field and temperature is mapped out. Remarkably, distinct from those reported in 3D systems, it can manifest two different antiferromagnetic skyrmion phases. In addition, through contacting with $Sc_2CO_2$, the AF-SkX in Janus monolayer $CrSi_2N_2As_2$ can be created or annihilated by ferroelectric switching. Our works promote further study on AFM skyrmion physics in 2D lattice.

**Methods**

Our first-principles calculations are performed based on density functional theory (DFT) as implemented in Vienna ab initio simulation package (VASP) [49,50]. The projector augmented wave (PAW) method is adopted to treat the ionic potential [51]. The Perdew-Burke-Ernzerhof (PBE) functional of generalized gradient approximation (GGA) is used for the exchange-correction interactions [52]. To describe well the strong correlations of 3d electrons, GGA+U method is adopted with effective Hubbard U＝3 eV for 3d electrons of Cr atom [53]. The plan-wave cutoff energy is set to 520 eV. The convergence criterion for force and energy are set to 0.001 eV/Å and 1 × 10$^{-6}$ eV, respectively. The Monkhorst-Pack k-point mesh of 13 × 13 × 1 is adopted to sample the 2D Brillouin zone. DFT-D3 method is employed for treating van der Waals interaction in heterostructures. Phonon spectra is obtained by PHONOPY code based on 2 × 2 × 1 supercell [54].

Using the magnetic parameters obtained from first-principles calculations, parallel tempering MC simulations with the Metropolis algorithm are carried out to get the energy minimum spin textures [55]. The spin textures are obtained based on a 120 × 120 × 1 supercell with 1 × 10$^5$ MC steps performed for each temperature (from 660 K gradually cooled down to the investigated low temperature).

**Results and Discussion**

**Fig. 1(a)** presents the crystal structure of Janus monolayer CrSi$_2$N$_2$As$_2$, which is derived from the prototype CrSi$_2$N$_4$ [56-58]. Each unit cell contains one Cr, two Si, two N and two As atoms, which are stacked in the sequence of N-Si-N-Cr-As-Si-As. It shows a hexagonal lattice with the space group *P6m2*. Therefore, the inversion symmetry is broken. The optimized lattice constant of Janus monolayer CrSi$_2$N$_2$As$_2$ is optimized to be 3.12 Å. Its phonon dispersions are shown in **Fig. S2(c)**, from which we can observe only a tiny imaginary frequency around the Γ point, confirming its stability.

The valence electronic configuration of Cr atom is *3d$^5$4s$^1$*. By coordinating with three N and three As atoms in monolayer CrSi$_2$N$_2$As$_2$, each Cr donates three electrons for bonding, resulting in the valence electronic configuration of *3d$^3$4s$^0$*. As shown in **Fig. 1(a)**, the six neighboring N/As atoms form a distorted trigonal prismatic geometry for Cr atom. In such a coordination environment, the $d$ orbitals of Cr atom roughly split into two groups, i.e., two high-lying $e_g$ and three low-lying $t_{2g}$ orbitals. The three left valence electrons of Cr atom half-fill the $t_{2g}$ orbitals, which would give rise to a magnetic moment of 3 $\mu_B$. Our first-principles calculations confirm that the magnetic moment distributed on Cr atom is 2.92 $\mu_B$. Therefore, monolayer CrSi$_2$N$_2$As$_2$ is a 2D magnetic material.

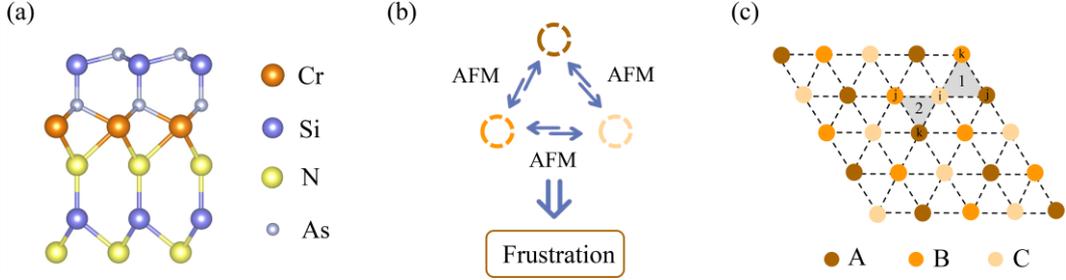

**Fig. 1.** (a) Crystal structure of Janus monolayer CrSi$_2$N$_2$As$_2$ from side view. (b) Geometric frustration from antiferromagnetic coupling in triangular lattice. (c) Diagram of three magnetic sublattices in triangular lattice.

To further investigate the magnetic properties, we construct the Heisenberg type spin Hamiltonian model of monolayer CrSi$_2$N$_2$As$_2$, which consists of nearest sites Heisenberg exchange interaction, DMI, single ion anisotropy (SIA) and Zeeman interaction. The effective Hamiltonian can be expressed as

$$H = -J \sum_{<i,j>} (\boldsymbol{S}_i \cdot \boldsymbol{S}_j) + \sum_{<i,j>} \boldsymbol{D}_{ij} \cdot (\boldsymbol{S}_i \times \boldsymbol{S}_j) - \lambda \sum_{<i,j>} (S_i^z \cdot S_j^z) - K \sum_i (S_i^z)^2 - h \sum_i S_i^z$$

Here, $\boldsymbol{S}_{i/j}$ is the unit vector ($|\boldsymbol{S}| = 1$) corresponding to local spin at $i/j^{th}$ Cr atom. $\boldsymbol{D}_{ij}$ and $K$ indicate the magnetic parameters characterizing DMI and SIA, respectively. The Heisenberg exchange interaction in the model is divided into two parts, i.e., isotropic and anisotropic terms. $J$ is the isotropic exchange interaction parameter ($J > 0$ indicates ferromagnetic coupling), and $\lambda$ is the anisotropic one. In the last term, $h$ indicates the strength of the external magnetic field.

The isotropic exchange interaction parameter $J$ is calculated to be -9.013 meV, which indicates the AFM neighboring interaction for monolayer CrSi$_2$N$_2$As$_2$. Such AFM neighboring interaction is related to its unique structure. According to the Goodenough-Anderson rules [59], FM coupling would dominate the exchange interaction between the neighboring magnetic moments when the bond angle is close to 90°; otherwise, AFM coupling would play a more important role. In monolayer CrSi$_2$N$_2$As$_2$, the bond angles of Cr-As-Cr and Cr-N-Cr are 78.13° and 100.58°, respectively, which are not close to 90° and thus give rise to the AFM neighboring interaction. As schematically shown in **Fig. 1(b)**, such AFM neighboring interaction in triangular lattice cannot result in anti-parallel arrangement for the magnetic moments of every two atoms at vertices of unit equilateral triangle. This would lead to the geometric frustration in monolayer CrSi$_2$N$_2$As$_2$. Namely, the Cr lattice presents a three-sublattice 120° structure [60], wherein the spins in each sublattice [colored in **Fig. 1(c)**] are parallelly arranged.

According to the Moriya's rule [61], the symmetry of triangular lattice enforces that $\boldsymbol{D}_{ij}$ for the nearest-neighboring Cr atoms is perpendicular to their bond. The DMI parameter $D$ is calculated to be 1.960 meV. Such a large value can be attributed to the combined effect of the large SOC strength within As atom and the large electronegativity difference between N and As atoms. This results in $D/J = 0.217$ for monolayer CrSi$_2$N$_2$As$_2$, which locates in the typical range for the formation of skyrmions (around 0.1 ~ 0.2) [62]. The SIA parameter $K$ is calculated to be -0.459 meV, indicating the in-plane magnetic anisotropy. While for the anisotropic exchange interaction parameter $\lambda$, it is estimated to be 0.122, which favors out-of-plane magnetic anisotropy. Obviously, $K$ plays a dominated role, estimating the in-plane magnetic anisotropy for monolayer CrSi$_2$N$_2$As$_2$. In this regard, monolayer CrSi$_2$N$_2$As$_2$ holds the possibility for hosting nontrivial topological spin structures.

**Table 1.** Lattice constants $a$, isotropic Heisenberg exchange interaction parameter $J$, in-plane DMI parameter $D$, anisotropic exchange interaction parameter $\lambda$, SIA parameter $K$ and magnetic moment $m$ on Cr atom for monolayer CrSi$_2$N$_2$As$_2$.

|  | $a$ (Å) | $J$ (meV) | $D$ (meV) | $\lambda$ (meV) | $K$ (meV) | $m$ ($\mu_B$) |
|---|---|---|---|---|---|---|
| CrSi$_2$N$_2$As$_2$ | 3.12 | -9.013 | 1.960 | 0.122 | -0.459 | 2.92 |

Based on the first-principles calculations parameterized Hamiltonian, we perform parallel tempering MC simulations to explore the spin configurations in monolayer CrSi$_2$N$_2$As$_2$. As displayed in **Fig. S4**, it exhibits a striped-like spin texture [referred to as spin spiral (SS) phase]. We then investigate its spin structures under external magnetic field and temperature. Here, we introduce the skyrmion number Q and chirality $\chi_L$ to detect the spin textures effectively. The skyrmion number Q and chirality $\chi_L$ are defined as [33]

$$Q = \frac{1}{4\pi}\sum_n q$$

$$\chi_L = \frac{1}{8\pi}\sum_n [\chi_{L,r_n^{(1)}} + \chi_{L,r_n^{(2)}}]$$

with
$$tan\frac{q}{2} = A_{L,r_n^{(1)}} \cdot \chi_{L,r_n^{(1)}} + A_{L,r_n^{(2)}} \cdot \chi_{L,r_n^{(2)}}.$$

Here, $A_{L,r_n^{(1,2)}} = 1/[1 + \mathbf{S}_i^{(1,2)} \cdot \mathbf{S}_j^{(1,2)} + \mathbf{S}_j^{(1,2)} \cdot \mathbf{S}_k^{(1,2)} + \mathbf{S}_k^{(1,2)} \cdot \mathbf{S}_i^{(1,2)}]$ is calculated in the $n^{th}$ unit equilateral triangle of the lattice, and the upper corner marks *1* and *2* represent two corner-shared triangles [highlighted in **Fig 1(c)**]. $\chi_{L,r_n^{(1,2)}} = \mathbf{S}_i^{(1,2)} \cdot (\mathbf{S}_j^{(1,2)} \times \mathbf{S}_k^{(1,2)})$ is the chirality localized in the unit equilateral triangle. $\mathbf{S}_i^{(1,2)}$, $\mathbf{S}_j^{(1,2)}$, $\mathbf{S}_k^{(1,2)}$ are the three spin vectors in the anticlockwise lattice. We calculate the Q and $\chi_L$ in three sublattices of the triangular lattice, wherein the spin vectors run in each sublattice. Also, we calculate total local chirality without sublattice distinction.

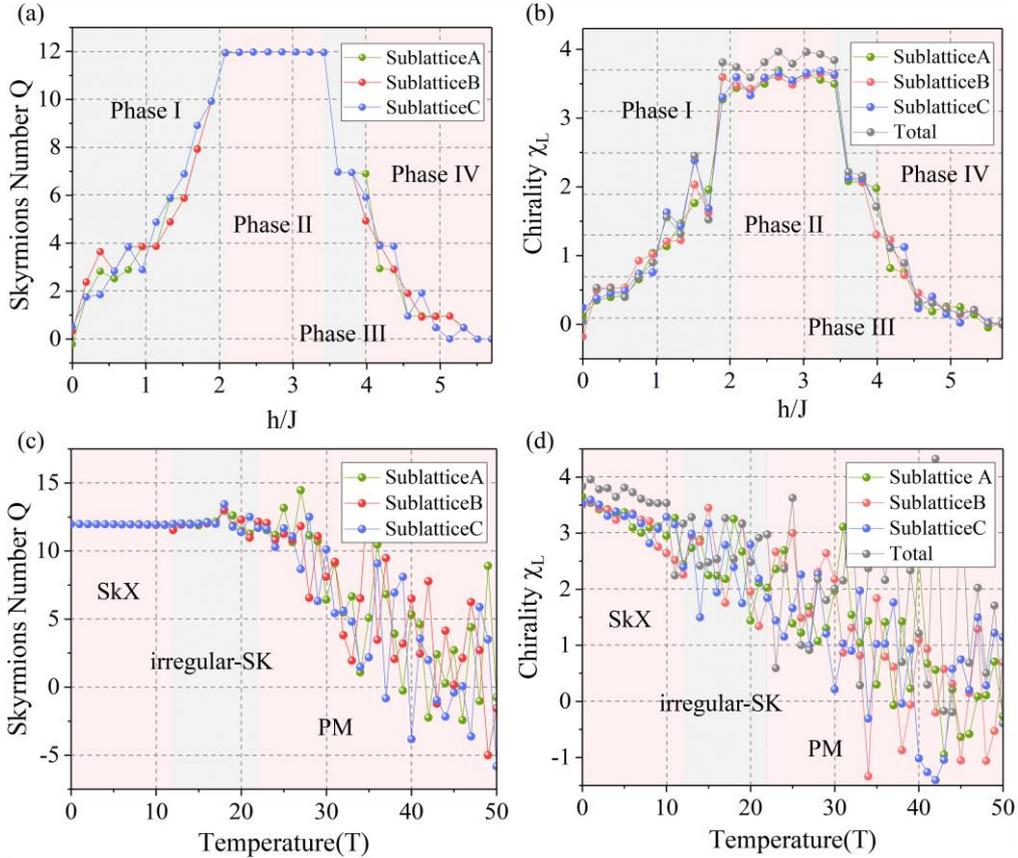

**Fig. 2.** Skyrmion number Q (a) and chirality $\chi_L$ (b) as a function of external magnetic field for monolayer $CrSi_2N_2As_2$. Skyrmion number Q (c) and chirality $\chi_L$ (d) as a function of temperature for monolayer $CrSi_2N_2As_2$ monolayer under $h/J = 2.85$.

**Fig. 2** presents the skyrmion number Q and chirality $\chi_L$ of monolayer $CrSi_2N_2As_2$ as a function of the external magnetic field. It can be seen that both Q and $\chi_L$ rise gradually under increasing external magnetic field $h/J$ from 0 to 2.08. Such continuous change suggests no phase transition and thus phase I is the SS phase. By further increasing $h/J$ to 3.42, intriguingly, Q of the three sublattices share the same values and remain unchanged at a platform, and $\chi_L$ are also stable with a slight fluctuation. These characters indicates that the topologically protected AF-SkX phase probably occurs. Upon increasing $h/J$ from 3.42 to 3.99, Q of the three sublattices jump to another platform and preserve unchanged, and $\chi_L$ remain stable with a slight fluctuation as well, suggesting another possible AF-SkX phase. Therefore, the monolayer $CrSi_2N_2As_2$ might exhibit two AF-SkX phases. When further increasing $h/J$ larger than 3.99, Q of the three sublattices are becoming different and gradually reduce to zero. This process corresponds to that the AF-SkX phase becomes broken and evaluates to SS phase finally.

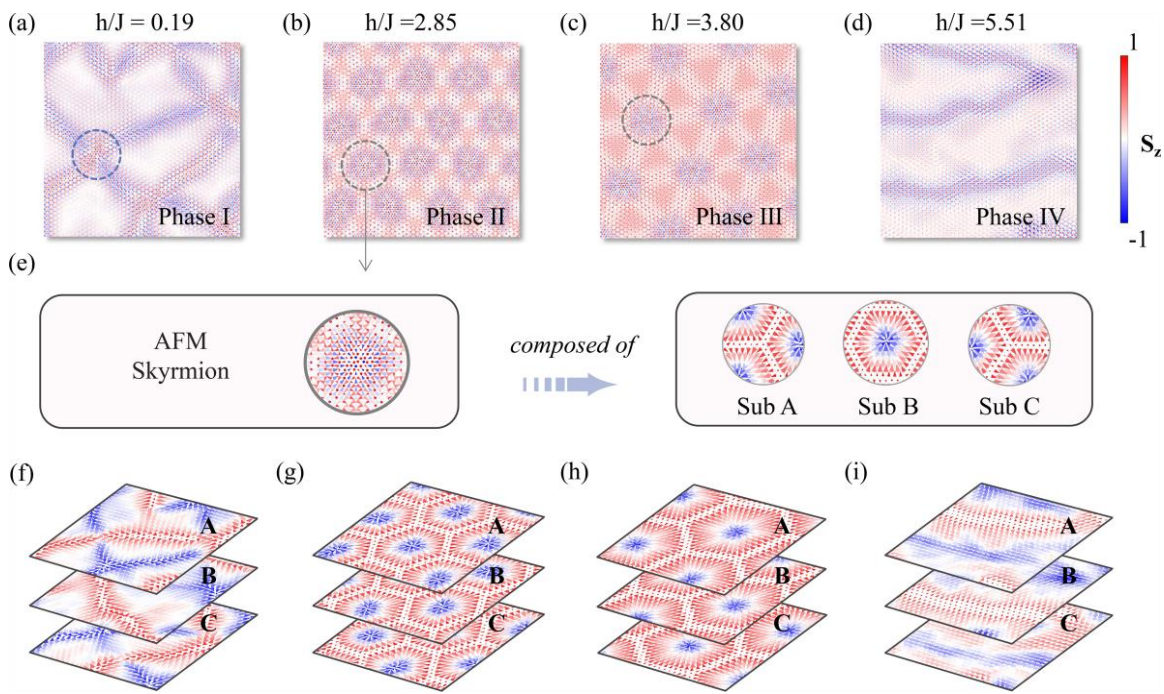

**Fig. 3.** Specific spin textures in triangular lattice and sublattices of monolayer $CrSi_2N_2As_2$ under (a, f) phase I, (b, g) phase II, (c, h) phase III, and (d, i) phase IV. (e) Spin patterns of isolated AFM skyrmion (left panel) and three FM skyrmions in sublattice A, B and C (right panel).

To get the specific spin patterns of monolayer $CrSi_2N_2As_2$ under external magnetic field, taking $h/J = 0.19$, 2.85, 3.80 and 5.51 as examples of these four phases, we show their spin structures in

**Fig. 3**. For more examples, please refer to **Fig. S4**. Under $h/J$ = 0.19, the spin structure of monolayer $CrSi_2N_2As_2$ is roughly similar with the case under $h/J$ = 0, displaying AFM SS phase [**Fig. 3(a)**]. As shown in **Fig. 3(f)**, such AFM SS phase can be considered as the interpenetration of the FM SS configurations in its three sublattices. From the spin texture shown in **Fig. 3(a)**, no skyrmion can be observed; however, the Q is calculated to be nonzero. Such nonzero Q is correlated to the vortices formed at the intersections of different stripes as Q is characterized by the product of vortex and polarity. Meanwhile, as shown in **Fig. S4**, with rising $h/J$ in phase I, the number of intersections increases and thus the Q increases.

For the case under $h/J$ = 2.85, as displayed in **Fig. 3(b)**, a regular array of AFM skyrmions is observed, giving rise to the intriguing AF-SkX phase. The AF-SkX is interpenetrated by the FM skyrmion crystals of its three sublattices, see **Fig. 3(g)**. For these three FM skyrmion crystals, the skyrmions are presented in the same size and density, and they can be converted to each other under a translation operation. From **Fig. 3(e)**, it can be seen that each AFM skyrmion in AF-SkX is formed by a FM skyrmion from one sublattice and two striped-like textures from the other two sublattices. Therefore, phase II is the topologically protected AF-SkX phase. **Fig. 3(c)** and **(h)** present the spin structure of monolayer $CrSi_2N_2As_2$ under $h/J$ = 3.80. Obviously, it exhibits a similar topologically protected spin texture, namely, AF-SkX phase. As compared with the case under $h/J$ = 2.85, the Q of monolayer $CrSi_2N_2As_2$ under $h/J$ = 3.80 is smaller, and thus its skyrmions are presented in a smaller density. Accordingly, monolayer $CrSi_2N_2As_2$ hosts two topologically protected AF-SkX phases. By increasing $h/J$ larger than 3.99, as shown in **Fig. S4**, the array of AFM skyrmions in monolayer $CrSi_2N_2As_2$ begins to be irregular, leading to the broken AF-SkX phase. Subsequently, the AFM skyrmions disappear and the sparse stripes appear, which yields to the SS phase [**Fig. 3(d)** and **(i)**].

**Table 2.** Lattice constants $a$, isotropic Heisenberg exchange interaction parameter $J$, in-plane DMI parameter $D$, anisotropic exchange interaction parameter $\lambda$, SIA parameter $K$ and magnetic moment $m$ on Cr atom for $CrSi_2N_2As_2/Sc_2CO_2$ heterobilayer under P+ and P-.

|     | $a$ (Å) | $J$ (meV) | $D$ (meV) | $\lambda$ (meV) | $K$ (meV) | $m$ ($\mu_B$) |
| --- | --- | --- | --- | --- | --- | --- |
| P+ | 3.29 | -3.061 | 1.355 | 0.080 | -0.475 | 3.03 |
| P- | 3.29 | -5.744 | 1.439 | 0.100 | -0.450 | 3.03 |

We then take the case under $h/J = 2.85$ as an example to investigate the influence of temperature on the nontrivial topological spin texture in monolayer $CrSi_2N_2As_2$. As shown in **Fig. 2(c)**, with increasing the temperature from 0 to 12 K, Q are almost unchanged, and the difference in Q among the three sublattices are negligible. While for $\chi_L$, although they have noticeable decrease, the differences among three sublattices are tiny; see **Fig. 2(d)**. These features indicate that AF-SkX phase in monolayer $CrSi_2N_2As_2$ can be preserved with the temperature reaching up to ~12 K. As rising of temperature larger than 12 K, both Q and $\chi_L$ begin to decrease with oscillation [**Fig. 2(c)** and **(d)**], which corresponds to the array of AFM skyrmions being to be irregular [called irregular skyrmion (irregular-SK) phase]. When the temperature is larger than ~22 K, the oscillations are chaotic and the amplitudes are huge, suggesting that the regular AF-SkX in monolayer $CrSi_2N_2As_2$ is annihilated thoroughly and paramagnetism (PM) phase appears.

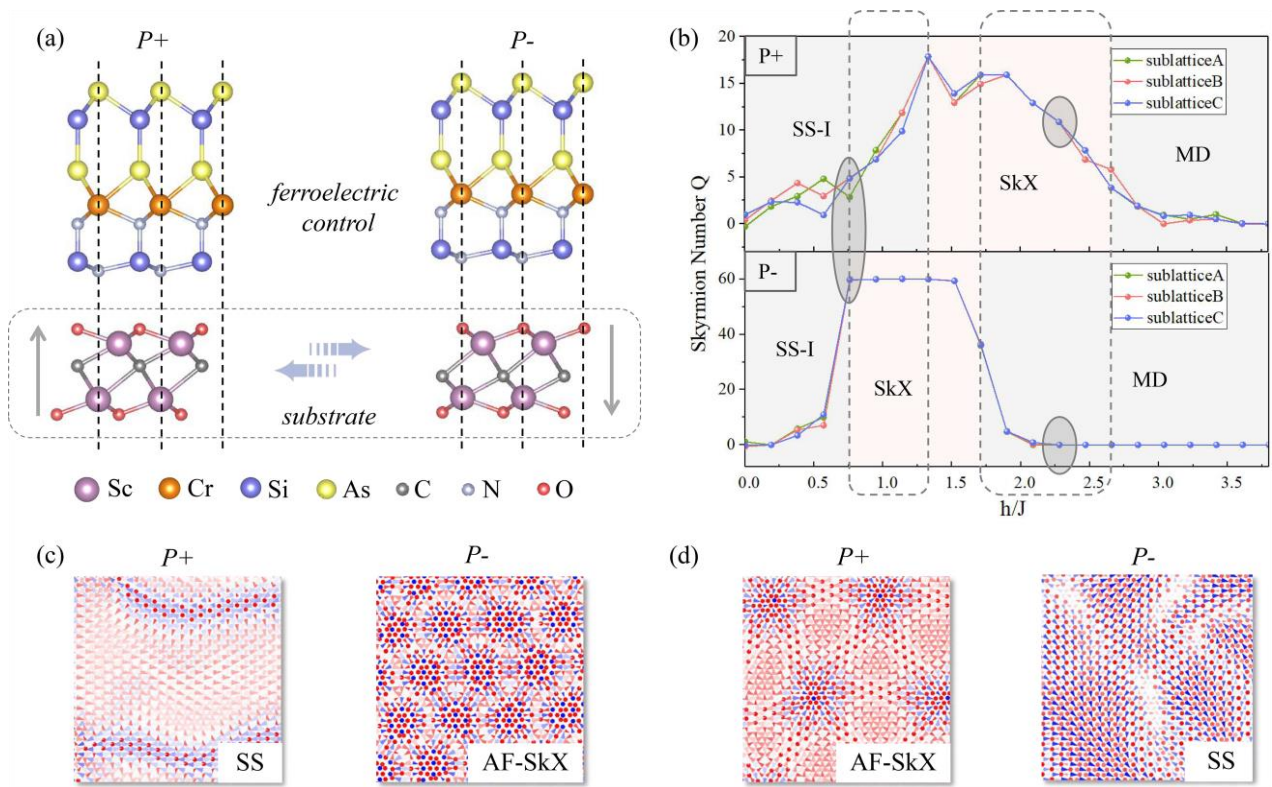

**Fig. 4**. (a) Crystal structures of heterobilayer $CrSi_2N_2As_2/Sc_2CO_2$ under P+ (left panel) and P- (right panel). (b) Skyrmion number Q as function of $h/J$ for P+ and P-. Specific spin patterns for P+ and P- with (c) $h/J = 0.76$ and (d) $h/J = 2.27$.

Having estimated the compelling topological spin structure in monolayer $CrSi_2N_2As_2$, we next propose a mechanism of coupling it with ferroelectricity for realizing electrical control of AF-SkX.

Here, monolayer $Sc_2CO_2$ is selected as the ferroelectric substrate because it exhibits a small lattice mismatch with monolayer $CrSi_2N_2As_2$. Given the switchable polarization of $Sc_2CO_2$, there are two types of heterobilayer $CrSi_2N_2As_2/Sc_2CO_2$: one with the electric polarization pointing along +z direction (P+), and the other with electric polarization pointing along -z direction (P-). **Fig. 4(a)** shows the crystal structures of heterobilayer $CrSi_2N_2As_2/Sc_2CO_2$. The corresponding magnetic parameters of P+ and P- are summarized in **Table 2**. It can be seen that as compared with the free-standing case, the absolute values of $J$ and $D$ both reduce significantly in P+ and P-. And the magnetic parameters of P+ and P- are different, which indicates that they have different topological spin structures.

**Fig. 4(b)** shows the skyrmion number Q of P+ and P- as a function of $h/J$. For P+, the SS phase is stable under $h/J = 0 \sim 1.33$, and transfers to the protected topologically AF-SkX phase under $h/J = 1.33 \sim 2.66$. By further increasing $h/J$ larger than 2.66, the AF-SkX is deformed and SS phase becomes stable again. Such variation of skyrmion number Q under $h/J$ is shared by P-, but the corresponding critical points are different, i.e., $h/J = 0.76$ and 1.70. Obviously, under $h/J = 0.76 \sim 1.33$ and $1.70 \sim 2.66$, the creation and annihilation of AF-SkX phase can be achieved by ferroelectric polarization switching. Here, taking $h/J = 0.76$ and 1.70 as examples of these two regions, we display their spin structures in **Fig. 4(c)** and **(d)**. It can be seen that under $h/J = 0.76$ (1.70), SS (AF-SkX) spin pattern appears in P+, which can be readily reversed to the AF-SkX (SS) phase upon switching the ferroelectric polarization, thus realizing the ferroelectrically tunable AF-SkX in monolayer $CrSi_2N_2As_2$.

At last, we discuss the strain effect on the magnetic parameters as well as topological spin structures of monolayer $CrSi_2N_2As_2$. As shown in **Fig. S7**, with increasing strain, $J$ increases monotonically, while $D$ decrease monotonically, leading to the monotonic decrease of $D/J$. This indicates that the topological spin structures of monolayer $CrSi_2N_2As_2$ would also vary with strain. **Fig. S8** shows the specific spin patterns of monolayer $CrSi_2N_2As_2$ under different strain and $h/J$. We can see that the AF-SkX phase appears under $h/J = 0.76$, 1.33 and 2.66, respectively, for the cases of 3%, 1% and -1% strain. While for the case of -3% strain, the AF-SkX phase is not observed within the considered $h/J$. By comparing with the pure case, it is interesting to note that strain can be used to significantly reduce the required $h/J$ for realizing the AF-SkX phase. Moreover, as shown in **Fig. S8**, along with increasing strain, the size of AFM skyrmions in AF-SkX is reduced, which is in favor for practical applications.

## Conclusion

In summary, using first-principles calculations and MC simulations, we demonstrate the existence of AF-SkX phase in 2D lattice of monolayer $CrSi_2N_2As_2$. The skyrmion physics in monolayer $CrSi_2N_2As_2$ is correlated to its large DMI and the geometric frustration of its triangular lattice. The evaluation of topological spin textures in monolayer $CrSi_2N_2As_2$ with *h/J* and temperature are comprehensively investigated. Distinct from those reported in 3D systems, two different antiferromagnetic skyrmion phases can be manifested in monolayer $CrSi_2N_2As_2$. Furthermore, with interfacing to monolayer $Sc_2CO_2$, the creation and annihilation of AF-SkX in monolayer $CrSi_2N_2As_2$ are achieved by ferroelectric polarization switching. The explored phenomena and insights greatly enrich the AFM skyrmion physics in 2D lattice.

## ACKNOWLEDGEMENT

This work is supported by the National Natural Science Foundation of China (No. 12074217), Shandong Provincial Natural Science Foundation (Nos. ZR2019QA011 and ZR2019MEM013), Shandong Provincial Key Research and Development Program (Major Scientific and Technological Innovation Project) (No. 2019JZZY010302), Shandong Provincial Key Research and Development Program (No. 2019RKE27004), Shandong Provincial Science Foundation for Excellent Young Scholars (No. ZR2020YQ04), Qilu Young Scholar Program of Shandong University, and Taishan Scholar Program of Shandong Province.

## COMPETING INTERESTS

The authors declare no competing interests.

## DATA AVAILABILITY

The authors declare that the data supporting the findings of this study are available within the paper and its supplementary information files.